\documentclass[article,showpacs,showkeys,superscriptaddress]{revtex4}

\usepackage{amsmath}
\usepackage{bbold}
\usepackage{amsfonts}
\usepackage{amssymb}
\usepackage{pbsi}
\usepackage[T1]{fontenc}
\usepackage{xcolor}
\usepackage{graphicx}
\usepackage{subfig}
\usepackage{float}
\usepackage{color}
\usepackage{hyperref}
\hypersetup{colorlinks = true,
	linkcolor = red,
	anchorcolor = blue,
	citecolor = blue,
	filecolor = blue,
	urlcolor = blue}

\begin{document}

\title{Finite Orbital Angular momentum Bessel beams propagating along light-cone coordinates }

\author{Felipe A. Asenjo}
\email{felipe.asenjo@uai.cl}
\affiliation{Facultad de Ingenier\'ia y Ciencias,
Universidad Adolfo Ib\'a\~nez, Santiago 7491169, Chile.}
\author{Swadesh M. Mahajan}
\email{mahajan@mail.utexas.edu}
\affiliation{Institute for Fusion Studies, The University of Texas at Austin, Texas 78712, USA.}

\date{\today}

\begin{abstract}
New solutions for Bessel electromagnetic beams, propagating along the light cones, are investigated. Of the variety of structures possible in the light cone variables, the one involving a product of Airy functions is discussed in detail. This class of solutions, representing an asymmetry on the light-cone coordinates dependence, is a non-trivial extension to the usual plane wave solutions. We also explore the conditions under which these solutions will carry finite orbital angular momentum density. 
\end{abstract}

\pacs{}

\keywords{}

\maketitle

\section{Introduction}

The standard solutions for electromagnetic waves in vacuum are plane waves propagating at the speed of light along any direction; these are the simplest elementary solutions of the wave equation emerging from Maxwell equations. 
One must wonder if the vacuum wave equation is capable of supporting solutions that are more complex and, hopefully, more interesting. One of these known solutions is the so called Bessel beam, that propagates along, let us say, the $z$-direction, while has a Bessel function behavior in the transverse spatial $x$-$y$ (more explicitly in the $r$-$\theta$) plane. Such electromagnetic wavepackets, propagating in the form of plane waves along the $z$-direction, have been extensively reported (see for instance Refs.~\cite{Efremidis,PHillion,pasi,mugnai,sesha,Colin,yurkin}). A highly interesting characteristic of the propagating Bessel beams is that they are endowed with an intrinsic orbital angular momentum (OAM)  (some examples are in Ref.~\cite{litvin,Dorrah,oscar,kot,alexa,wangjian}).

This work is devoted to investigating new propagating solutions accessible to the Bessel beam class; they constitute a novel departure from of previous known wavepackets in that the new solutions could  propagate along the two light-cone coordinates  given by $\eta=z-t $ and $\xi= z+t$, in the $z$-$t$ plane.  We will demonstrate that these solutions  have a wide variety of structures in the light-cone coordinates, allowing us to construct these new Bessel beams with OAM properties.


The particular solution we will analyze in detail, is algebraically a multiplication of Airy functions (with arguments that are complicated combinations of  $\eta$ and $\xi$), showing  an asymmetry between the propagation in the vicinity of two light-cone coordinates. These wavepackets are structurally quite complex and are not just a simple change of variables in the plane wave solution. In addition, due to the transversal structured extension, the solutions do not propagate at the speed of light.

In the $\eta$-$\xi$ plane, several other solutions of the same family, are also found in terms of a multiplication of   Parabolic Cylinder, Mathieu and Modified Bessel functions. These solutions, though equally interesting, are not discussed now; they are displayed in Appendix A.

The light cone-Bessel beams are studied by starting from the vacuum Maxwell equations $\partial_t{\bf E}=\nabla\times{\bf B}$ and $\partial_t{\bf B}=-\nabla\times{\bf E}$, for  ${\bf E}$ (${\bf B}$) the electric (magnetic) field \cite{Landau}.  
These fields can be described in terms of a scalar ($\Phi$) and vector (${\bf A}$) potentials, ${\bf E}=-\partial_t {\bf A}-\nabla\Phi$ and ${\bf B}=\nabla\times {\bf A}$, such that their dynamics can be found 
through the wave equation
\begin{equation}
    \left(\frac{\partial^2}{\partial t^2}-\nabla^2\right){\bf A}=0\, ,
    \label{waveeq2bb}
\end{equation}
under the condition $\partial_t\Phi+\nabla\cdot{\bf A}=0$ (Lorentz gauge).
We will now manipulate this very simple looking equation to extract rather complicated solutions. From all the possible choices, we will explore separable solutions of the form \begin{eqnarray}
    {\bf A}(t,{\bf x})=a(t,z) {\bf u}(x,y)=  a(t,z) {\bf u}(r,\phi),
    \label{ansatz1}
\end{eqnarray}
that propagates in the $z$-$t$ plane, and have a spatial structure (to be determined) in the transverse $x$-$y$ plane, which is the $r$-$\phi$ plane in cylindrical coordinates. Here, $a$ is a scalar function while ${\bf u}$ contains the direction of the potential. For this from of ${\bf A}$ , the magnetic and electric field can be readily obtained as
\begin{eqnarray}
    {\bf B}&=&\nabla\times(a\,  {\bf u})\, ,\nonumber\\
    {\bf E}&=&-\frac{\partial a}{\partial t}\, {\bf u}+\nabla\left(\nabla\cdot(b\, {\bf u})\right), \quad \mbox{with}\quad \frac{\partial b}{\partial t}=a\, .    \label{electricmagneticgen}
\end{eqnarray}

Equations~\eqref{waveeq2bb} and \eqref{electricmagneticgen} are the starting point of our enquiry. Because of the separability assumption, \eqref{waveeq2bb} can be split as ($\lambda$ is an arbitrary separable constant)
\begin{equation}
   \frac{1}{a} \left(\frac{\partial^2}{\partial t^2}-\frac{\partial^2}{\partial z^2}\right)a= \frac{\nabla^2 \bf{u}}{
   \bf{u}}=-\lambda^2
    \label{waveeq2bb1}
 \end{equation}
implying that these independent equations
\begin{equation}
    \left(\frac{\partial^2}{\partial t^2}-\frac{\partial^2}{\partial z^2}+\lambda^2 \right)a=0\, 
    \label{waveeq21}
  \end{equation}
and 
 \begin{eqnarray}
     \hat{e}_r\cdot \nabla^2 {\bf u}+\lambda^2{u_r}=0 , \quad   \hat{e}_{\phi}\cdot\nabla^2 {\bf u}+\lambda^2{u_{\phi}}=0 
     \label{transverseEq}
 \end{eqnarray}
must be solved to construct the final solution for $a=a(t,z)$ and ${\bf{u}}= \hat{e}_r u_r+ \hat{e}_{\phi} u_{\phi}$ (although $u_z(r,\phi)$ could be easily retained, we will deal with only the transverse components of ${\bf{u}}$). Naturally the Bessel character of the wavepacket will emerge through $\bf{u}$. However, we will first calculate the light cone structures (in $\eta$ and $\xi$)  that can be harbored by \eqref{waveeq21} for this Bessel beam. 
%
%

\section{Bessel light beams in light-cone coordinates}
\label{sec1}

Notice that \eqref{waveeq21} mimics an electromagnetic wave with inertia as if it was traveling in a dielectric medium. This effective inertia for this traveling in vacuum is of course, is induced by nonzero transversal dependence, described by ${\bf u}$.
We now look for particular solutions of Eq.~\eqref{waveeq21} that are separable in light-cone coordinates variables, $\eta=z-t$ and $\xi=z+t$, 
\begin{equation}
a(t,z)=a(\rho,\chi)=f(\rho)g(\chi)\, , 
\end{equation}
where now $f$ and $g$ are arbitrary functions of $\rho=\rho(\eta,\xi)$ and $\chi=\chi(\eta,\xi)$. The preceding {\it ansatz} converts the  wave equation \eqref{waveeq21} into
\begin{eqnarray}
\left[\partial_\eta\rho\partial_\xi\rho\frac{\partial^2}{\partial\rho^2}+\left(\partial_\eta\rho\partial_\xi\chi+\partial_\xi\rho\partial_\eta\chi\right)\frac{\partial^2}{\partial\chi\partial\rho}+\partial_\eta\chi\partial_\xi\chi\frac{\partial^2}{\partial\chi^2}\right]fg-\frac{\lambda^2}{4}fg=0\, .
\label{waveeq3}
\end{eqnarray}
The choice
\begin{eqnarray}    \rho(\eta,\xi)&=&\Theta(\eta)+\Phi(\xi)\, ,\nonumber\\
    \chi(\eta,\xi)&=&\Theta(\eta)-\Phi(\xi)\, ,
    \label{ansatzimportant}
\end{eqnarray}
works miraculous simplification reducing Eq.~\eqref{waveeq3} to
\begin{equation}   
\partial_\eta\Theta\partial_\xi\Phi\left(\frac{\partial_\rho^2 f}{f}-\frac{\partial_\chi^2 g}{g}\right) -\frac{\lambda^2}{4}=0\, .
   \label{waveeq4}
\end{equation}

It is quite remarkable that \eqref{waveeq4} has the potential of describing a variety of wave packets as functions of the light-cone coordinates $\eta$ and $\xi$ \cite{lightasenjo}. Since all these could have a radially Bessel behavior, different functional forms of $f$ and $g$ will each generate an independent Bessel beam wavepacket.  

To make sure that we are on track, let us first solve Eq.~\eqref{waveeq4} to recover the usual plane wave solutions in the direction of propagation. This is just the solution usually associated with Bessel beams. The exponential functions $f=\exp(i\alpha\rho)$ and $g=\exp(i\beta\chi)$ ($\alpha\neq \beta$) converts Eq.~\eqref{waveeq4} to
\begin{equation}   
\partial_\eta\Theta\partial_\xi\Phi-\frac{\lambda^2}{4(\beta^2-\alpha^2)}=0\, .
\end{equation}
leading to  $\Theta=\lambda\eta/(2\sqrt{\beta^2-\alpha^2})$, $\Phi=\lambda\xi/(2\sqrt{\beta^2-\alpha^2})$, and $f(\rho)g(\chi)=\exp[i\lambda(\alpha z-\beta t)/\sqrt{\beta^2-\alpha^2}]$; this is, indeed, the conventional plane wave solution. The requirement $\beta>\alpha$ implies that the electromagnetic wave cannot propagate at the speed of light due to the effective inertia induced by $\lambda$.  


\subsection{Double Airy-Bessel beam}
\label{douairy}

Of all the solutions accessible to Eq.~\eqref{waveeq4} (the defining equation for the light cone structure of Bessel Beams)
we will now focus on, arguably, the simplest non-trivial one where $f$ and $g$ are both Airy functions (several others are shown in Appendix A) obeying ${\partial_\rho^2 f}=\rho {f}$ and ${\partial_\chi^2 g}=\chi {g}$. The obvious manipulation, ${\partial_\rho^2 f}/{f}-{\partial_\chi^2 g}/{g}=\rho-\chi=2\Phi$, reduces Eq.~\eqref{waveeq4} to
\begin{equation}
   \partial_\eta\Theta\, \partial_\xi\Phi^2-\frac{\lambda^2}{4}=0\, ,
   \label{waveeq6}
\end{equation}
that, in turn, fixes the ($\eta$, $\xi$) dependence of $\Theta$ and $\Phi$ as
 \begin{eqnarray}
\Theta=\frac{\eta}{\alpha^2}\, , \quad \Phi=\frac{\alpha \lambda\sqrt{\xi}}{2}\, ,
\end{eqnarray}
in terms of an arbitrary parameter $\alpha$. This, then, fixes the arguments ($\rho$ and $\chi$) of the Airy solutions. For other choices (displayed in Appendix A) the arguments of the relevant special functions will be different.

Due to the linearity of the defining equations, both functions $f$ and $g$ could be either of the Airy functions $\mbox{Ai}$ or $\mbox{Bi}$ (or some combination thereof). The simplest combination, displayed 
in terms of space and time coordinates, perhaps, is  \cite{otk,lightasenjo}
\begin{eqnarray}
a(t,z)&=&\mbox{Ai}\left(\frac{z-t}{\alpha^2}+\frac{\alpha\lambda}{2}\sqrt{z+t} \right)\mbox{Ai}\left(\frac{z-t}{\alpha^2}-\frac{\alpha\lambda}{2}\sqrt{z+t} \right)\, .
\label{Azdoubleairy}
\end{eqnarray}
Notice that this double ${\mbox{Ai}}$ electromagnetic wavepacket is asymmetric between the two light-cone coordinates, and therefore, is not a trivial extension of the plane wave solution. For example, when $\alpha$ becomes small, the wavepacket travels in a direction very close to the light-cone coordinate $\eta$. It is important to remark that this Double ${\mbox{Ai}}$ solution is an exact solution, and not a paraxial limit of Maxwell equations. 

The second function $b(t,z)$ for needed for evaluating the fields, may be readily calculated,
\begin{eqnarray}
b(t,z)&=&\int^t dt'\, \mbox{Ai}\left(\frac{t'-z}{\alpha^2}+\frac{\alpha\lambda}{2}\sqrt{z+t'} \right)\mbox{Ai}\left(\frac{t'-z}{\alpha^2}-\frac{\alpha\lambda}{2}\sqrt{z+t'} \right)\, .
\label{Azdoubleairy3b}
\end{eqnarray}

More complex solutions of Eq.~\eqref{waveeq6} can be fashioned by invoking both $\mbox{Ai}$ and $\mbox{Bi}$ (see Appendix B).  
At this juncture, it is important to remember that we are dealing with a linear system, and  distinct solutions may be constructed by superposition.

\subsection{Poynting vector  and Orbital angular momentum.}

How do these double Airy solutions differ from conventional plane wave solutions? To answer this question, we must delve into the details of the associated energy/momentum flow, that is, we must examine 
 the Poynting vector ${\bf S}=({\bf E}^*\times{\bf B}+{\bf E}\times{\bf B}^*)/2$ (where $^*$ stands for complex conjugate) associated with the  wave packet. 
From the relation \eqref{electricmagneticgen}, we notice that   the Poynting vector can be fully evaluated in terms of $a$ (and $b$), and the transverse part, that we will soon show, is evaluated in terms of Bessel functions.
 
Since  the electric and magnetic fields have components in all directions,
the Poynting vector also has nonzero components in the $r$-$\phi$ plane in addition to the $z$ component (typical of the conventional plane wave).  The energy-momentum flow, thus, is not limited to what would be direction of propagation for a plane wave. It is this transverse flow that causes sub luminal energy propagation of the Bessel beam. 

An interesting feature of Bessel beams lies in their ability to carry an intrinsic orbital angular momentum. For any wavepacket, the total angular momentum density can be calculated as ${\bf L}={\bf r}\times {\bf S}$ \cite{litvin}, where ${\bf r}$ is a position vector in the transversal plane to the propagation. In general, along the direction of propagation, the OAM density \cite{litvin} for the current wavepacket is 
\begin{equation}
    L_z={\hat e_{z}}\cdot{\bf L}= r S_{\phi}=\frac{r}{2}(E_z B_r^*+E_z^*B_r)-\frac{r}{2}(E_r B_z^*+E_r^*B_z)\, .
    \label{oamlz}
\end{equation}

In order to evaluate the OAM for the new Double Airy-Bessel beam, we need a solution for both components of Eq.~\eqref{transverseEq}. In cylindrical coordinates, they are readily found to be 
 \begin{equation}
   {\bf u}(r,\phi)=J_{l+1}\left(\lambda\, r\right) \left[\hat{e}_{r}\, \cos(l\phi)+ \hat{e}_{\phi}\, \sin(l\phi)\right]\, ,
\label{formu0}
\end{equation}
where $J_{l+1}$ is the Bessel function of order $(l+1)$. 
Since $a(t,z)$ and  ${\bf u}$ are real  so will be the electric and magnetic fields. The sought after  OAM, then, will be 
\begin{eqnarray}
    L_z=\frac{\lambda\, r}{2}\sin(2l\phi)\left( \frac{a\, b}{r}\, J^2_l(\lambda r)   -{\lambda a\, b}\, J_l(\lambda r)J_{l-1}(\lambda r)+\left(a\frac{\partial a}{\partial t}-  \frac{\partial a}{\partial z}\frac{\partial b}{\partial z}\right)J_{l}(\lambda r)J_{l+1}(\lambda r)\right)\, ,
    \label{OAMdensityz0}
\end{eqnarray}
and is non zero for all $l\neq 0$. It is usual that, in the OAM \eqref{OAMdensityz0}, the terms with derivatives of the propagating functions $a$ and $b$ are neglected under the assumption of paraxial approximation. But here, we display the complete solution as we are dealing with an exact electromagnetic wavepacket.

\section{Remarks}
\label{concl}

In this paper, we have proposed and examined new exact structures for the wavepackets of light traveling in vacuum along its light-cone coordinates, where its radial dependence is given in terms of Bessel functions. These Bessel beams can acquire a rich variety due to these new light-cone solutions. In here, we have examined in detail the Bessel beam with a double Airy structure.
One of the most remarkable feature of the investigated solution is that it exhibits an asymmetry in the propagation on the two light-cone coordinates. This asymmetry stems from the ansatz 
\eqref{ansatzimportant}, which simplifies the wave equation \eqref{waveeq3}, makes it separable and hence solvable. The (perhaps intended) consequence  is that it creates a tunable preference on the direction of propagation; by properly select the arbitrary constant in the solutions, one can force the transverse Bessel beams to propagate almost purely along one of light-cone coordinates at a velocity arbitrarily close to the speed of light. 

Furthermore, this new solution comes to complement all the previous studies that produce electromagnetic wavepackets with OAM using simple plane waves.  The previously presented 
special class of solutions did impart new and different longitudinal structures for the propagation of Bessel beams, enriching the theoretical studies of these beams. By the same token, we expect that new wavepackets would be experimentally realizable in laboratory just like Bessel beams with simpler structures.

However, and differently to what occur with electromagnetic plane waves, the OAM \eqref{OAMdensityz0} of the Double Airy-Bessel beams has a fully four-dimensional structure; its time  dependence is totally non-trivially. This feature, inherited from the structured form of propagation defined by the product of Airy functions, is the very opposite of what occurs in the simple plane wave solution (in $z$ and $t$); in the latter,  OAM depends only on the transversal structure of the beam.

Finally, although we have explored only the double $\mbox{Ai}$ solution in detail, many others are listed  in the Appendix A. Also, the space of new wavepackets can be substantially increased by a different choice for the simplifying ansatz  in Eq.~\eqref{waveeq3}. The newer structure of Bessel beams will be explored in future work.

\begin{acknowledgements}
FAA thanks to FONDECYT grant No. 1230094 that partially supported this work. The work of SMM was supported by U.S. Department of Energy Grant No. DE-FG02-04ER-54742.
 \end{acknowledgements}

\appendix

\section{Other solutions along light-cone coordinates}
\label{solutions}

In here, we show how Eq.~\eqref{waveeq4} is solved by looking functions $f$ and $g$ with different dependence on the light-cone coordinates.

{\it Double Parabolic Cylinder mode}.
We can solve Eq.~\eqref{waveeq4} if both $f$ and $g$ satisfy the generalized Weber equation $W''(y)=(a y^2+b y-n)W(y)$, where $a$, $b$, and $n$ are constants. The solutions of this equation can be written in terms of
parabolic cylinder function $D_{p}(\sqrt{2}a^{1/4}y+b/\sqrt{2 a^{3/2}})$, with $p=n/(2\sqrt{a})+b^2/(8a^{3/2})-1/2$.
In this case, ${\partial_\rho^2 f}/{f}-{\partial_\chi^2 g}/{g}=2\Phi(2a\Theta+b)$, and  Eq.~\eqref{waveeq4} becomes
$\partial_\eta\left(a\Theta^2+b\Theta\right)\, \partial_\xi\Phi^2={\lambda^2}/{4}$.
The solution of this equation is given  by $\Theta=-b/(2a)+\sqrt{b^2/(4a^2)+\eta/(a\alpha^2)}$, and $\Phi=\alpha\, \lambda\sqrt{\xi}/2$, where $\alpha$ is arbitrary.

{\it Double Mathieu  mode}.
Other solution of Eq.~\eqref{waveeq4} can be obtained if both  $f$ and $g$ satisfy the Mathieu equation $M''(y)+(a-b\cos(y))M(y)=0$, with solutions $M_{a,b}(y)$ \cite{bender} (where $a$ and $b$ are constants).
In this case, 
 ${\partial_\rho^2 f}/{f}-{\partial_\chi^2 g}/{g}=a-b\cos\rho-(a-b\cos\chi)=2b\sin\Theta\sin\Phi$. Therefore, Eq.~\eqref{waveeq4} becomes $\partial_\eta(\cos\Theta) \,\partial_\xi(\cos\Phi)={\lambda^2}/({8b})$,
that has solutions $\cos\Theta=\alpha\eta$, and $\cos\Phi=\lambda^2 \xi/(8b\alpha)$, for an arbitrary $\alpha$.

{\it Double Modified Bessel mode.} In this case, we have that $\partial_\rho^2f=a\, e^{b\rho} f$, and 
$\partial_\chi^2g=a\, e^{b\chi} g$, where $a$ and $b$ are constants, whose solution is $f(\rho)=K_0(2\sqrt{a} e^{b\rho/2}/b)$ (and similar solution for $g$). Here,  $K_0$ is the modified Bessel function of order 0.
Therefore, $\partial_\rho^2f/f-\partial_\chi^2g/g=2a e^{b\Theta}\sinh(b\Phi)$, and Eq.~\eqref{waveeq4} becomes $\partial_\eta(e^{b\Theta}) \,\partial_\xi(\cosh(b\Phi))={\lambda^2b^2}/({8a})$.
The solutions are $e^{b\Theta}=\alpha\eta$, and $\cosh(b\Phi)=\lambda^2 b^2\xi/(8a\alpha)$, for arbitrary $\alpha$.

\section{Other types of Double Airy-Bessel beams}

We present two different solutions of Eq.~\eqref{waveeq6} in terms of Airy functions.

First, let us choose a purely imaginary $\alpha= i \beta$ (with  $\beta$ real) and construct  the  combination 
\begin{eqnarray}
a(t,z)&=&\mbox{Ai}\left(\frac{t-z}{\beta^2}+\frac{i\beta\lambda}{2}\sqrt{z+t} \right)\mbox{Bi}\left(\frac{t-z}{\beta^2}-\frac{i\beta\lambda}{2}\sqrt{z+t} \right)\, .
\label{Azdoubleairy2ap1}
\end{eqnarray}
This propagating solution is complex (as usual plane wave solution is).

 Another different solution can be found  by looking Airy solutions of Eq.~\eqref{waveeq6} 
with an extra arbitrary constant freedom  that was not used in constructing Eq.~\eqref{Azdoubleairy}. The defining equations for $f$ and $g$ [solving Eq.~\eqref{waveeq6}] could be written as ${\partial_\rho^2 f}=(\rho+i \beta){f}$, and ${\partial_\chi^2 g}=(\chi+i\beta) {g}$,
with  a real arbitrary constant $\beta$. The double ${\mbox{Ai}}$ solution will, then, have the form
\begin{eqnarray}
a(t,z)&=&\mbox{Ai}\left(\frac{z-t}{\alpha^2}+\frac{\alpha\lambda}{2}\sqrt{z+t} +i\beta\right)\mbox{Ai}\left(\frac{z-t}{\alpha^2}-\frac{\alpha\lambda}{2}\sqrt{z+t} +i\beta\right)\, ,
\label{Azdoubleairy2ap2}
\end{eqnarray}
with real $\alpha$. The inclusion of the $\beta$ term allow that $a$  (and $b$) be complex.  

In both previous cases \eqref{Azdoubleairy2ap1} and \eqref{Azdoubleairy2ap2}, the electromagnetic propagation will have OAM.
This is achievable, for instance, when the vector field ${\bf u}$ that solves Eq.~\eqref{transverseEq} is also required to be complex, 
${\bf u}(r,\phi)=J_{l-1}\left(\lambda\, r\right)\,e^{i l \phi} ( \hat{e}_{r}+i \hat{e}_{\phi})$. In this case,  OAM \eqref{oamlz} becomes
\begin{eqnarray}
    L_z=\frac{i l}{2}J_l^2(\lambda r) \left(a\, b^*-a^*\, b\right)+\frac{i \lambda r}{2}J_l(\lambda r)J_{l-1}(\lambda r)\left[\frac{\partial a}{\partial t}a^*-\frac{\partial a^*}{\partial t}a+\frac{\partial b^*}{\partial z}\frac{\partial a}{\partial z}-\frac{\partial b}{\partial z}\frac{\partial a^*}{\partial z}+\lambda^2(b a^*-ba^*)\right]\, .
    \label{generaloam}
\end{eqnarray}

\end{document}